\documentclass[a4paper,12pt, notitlepage]{article}

\usepackage[top=25mm,bottom=25mm,left=25mm,right=25mm]{geometry}

\usepackage{tikz}
\usepackage[compat=1.1.0]{tikz-feynman}

\usepackage{amsmath} 
\usepackage{graphicx}
\usepackage{epstopdf}
\usepackage{url}
\usepackage{setspace} 
\usepackage{titlesec}
\titleformat{\section}{\bfseries\large\scshape\filcenter}{\thesection}{1em}{}
\titleformat{\subsection}{\bfseries\normalsize\scshape\filcenter}{\thesubsection}{1em}{}

\newcommand{\captionfonts}{\footnotesize}
\renewcommand\thesection{\Roman{section}.}
\renewcommand\thesubsection{\Alph{subsection}.}

\makeatletter
\long\def\@makecaption#1#2{
  \vskip\abovecaptionskip
  \sbox\@tempboxa{{\captionfonts #1: #2}}%
  \ifdim \wd\@tempboxa >\hsize
    {\captionfonts #1: #2\par}
  \else
    \hbox to\hsize{\hfil\box\@tempboxa\hfil}%
  \fi
  \vskip\belowcaptionskip}

\renewcommand\p@subsection{\thesection}
    
\makeatother

\begin{document}
\title{\textbf{\large{Experimental Search For Interference Effects In qq At LHC 
Final Report}}}

\author{\normalsize{Arman Boroumand Naeini} \\
        \small\textit{
        Department of Physics, University of Warwick,
        Coventry CV4 7AL, United Kingdom}}
\date{\today}

\maketitle 
\vspace{-10mm}

\begin{abstract} 
\noindent
At present, the simulations for hadronic Z decay used at the Large Hadron Collider (LHC) neglect the effects of interference between the electroweak production of the hadronic Z boson and the Quantum Chromodynamics (QCD) production of $q\bar{q}$. Indeed, the diboson Monte Carlo samples are generated independently from QCD processes. The purpose of this project is to assess the validity of this simulation approach by studying these interference effects, comparing Monte Carlo samples to experimental data. In particular, regions of low transverse momenta are explored for the first time, revealing new challenges unique to these regions.  Three sigma evidence for ZZ $\rightarrow$ \emph{b}$\bar{b}\mu\bar{\mu}$ is found at low transverse momenta for the very first time.
However, the error on the mass peak is too large to provide any sufficient evidence for interference. As such, there is no evidence to suggest that the simulation strategies adopted at the LHC are inadequate.

\end{abstract}
\vspace{6mm}


\section{Introduction}
This final report is a bachelor's thesis which sets out to investigate the interference effects between the electroweak production of the hadronic Z boson and the Quantum Chromodynamics (QCD) background at the Large Hadron Collider (LHC) at CERN, Switzerland. Currently, these interference effects are not accounted for in simulations at the LHC. The aim of this project is to test the validity of this simulation strategy at low values of transverse momenta ($p_{T}$) by exploring the significance of this interference effect on the mass of the Z boson, paving the way for future research in this field. 

An early theoretical study~\cite{Paper_1} explored the interference effect in single and double-jet cross sections in \emph{pp} and \emph{p}$\bar{p}$ collisions at energies lower than 1 TeV. The paper finds that ``the interference terms strongly influence the cross section''~\cite{Paper_1}. A later paper~\cite{Paper_2} studied the W and Z resonance peaks in double-jet production in \emph{pp} collisions, stating that the ``QCD-electroweak interference effects substantially alter both the shape of the invariant mass distribution and the event rate and must be included if the W, Z-, two-jet rates are to be determined from the measured distribution''~\cite{Paper_2}. Working with collisions at 630 GeV and 1.8 TeV, the paper found that QCD-weak interference effects caused the mass peak of the hadronic vector bosons to decrease by about 0.3 GeV. It was also noted that this shift could be even greater when the experimental resolution of the detector is taken into consideration. The paper concluded that their results provide evidence that ``effects caused by the interference between electro-weak and QCD amplitudes must be taken into account when data are compared with theoretical predictions''~\cite{Paper_2}. The paper also clearly explains the difficulties in extracting a signal in such interactions and in finding a mass shift, this will be returned to at a later stage. Although the Z boson had been detected from leptonic decay in 1983 by the UA1 and UA2 experiments~\cite{Paper_3}, it was not until 1990 when the hadronic decay of the Z boson had been detected by UA2 experiments in double-jet decays~\cite{Paper_4}. The experiment used 4.66$pb^{-1}$ of $sp\bar{p}s$ data and their results assumed a 2.2 GeV shift in the experimental mass peak position due to interference, based on Ref.~\cite{Paper_2}. J. Pumplin~\cite{Paper_5} studied the double-jet hadronic decays of the vector bosons at Fermilab's hadron collider, Tevatron. Pumplin detailed a method to calibrate the detector and the QCD Monte Carlo (MC) simulations of the hadron showers using the mass of the Z and W boson, observing that interference effects result in a downward shift of about 0.35 GeV in the mass peak. The latest paper~\cite{Paper_6}, published in 2020, analysed the interference effects at the LHC, considering the hadronic decay of the Z and W boson as the signal and the QCD background. The paper explored ``the effect of this interference on the reconstructed peak positions and rates for several production modes'', considering boosted vector bosons and vector boson pairs. The paper concluded that although interference effects appear to be small for boosted vector bosons, experiments accessing lower transverse momenta could see ``much larger'' interference effects. This project aims to pick up from the conclusion of Ref. \cite{Paper_6}, with the aim of exploring these interference effects at lower transverse momenta at the LHC.

\section{Core Theoretical Background}
In this section, the key interactions and Feynman diagrams will be presented and interference will be discussed. It is noted that natural units are used throughout the report.

To study interference, the key focus of this project is on the following process:

\begin{equation}
q\bar{q} \rightarrow q\bar{q}
\label{equation7}
\end{equation}

\begin{figure}[b!]
\centering
\includegraphics[width=150mm]{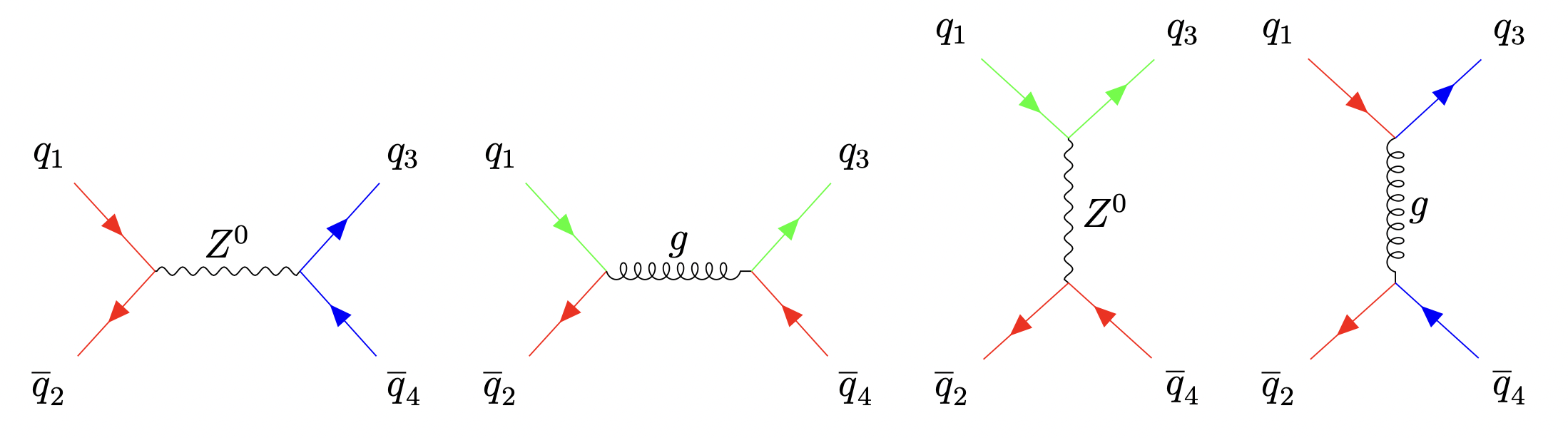}
\caption{ LO Feynman diagrams for \emph{q}$\bar{q}$ $\rightarrow$ \emph{q}$\bar{q}$. First two diagrams are the s-channel processes and second two are the t-channel processes. Also note that the Z can be replaced with a W boson or a photon.}
\label{fig: fig_1}
\end{figure}

Figure \ref{fig: fig_1} shows the Feynmann diagrams for the leading order (LO) electroweak and QCD processes for this interaction. The s-channel electroweak diagram is the only process which produces a resonant Z mass signature in the final \emph{q}$\bar{q}$ state through quark-antiquark annihilation and production. Here, it is importantly noted that this process is a colour singlet, that is, there is only one colour present in the final state. Interference only occurs between processes with identical input and output states. Therefore, the t-channel gluon QCD interaction contributes to the interference, but the s-channel QCD interaction does not since it contains more than one colour in the final state: it is a colour octet. The s-channel electroweak process is the signal that is being looked for. The background comprises every other interaction that produces a double-jet, therefore the background has large contributions from QCD processes with gluons in the initial and final states. As highlighted by Ref. \cite{Paper_2}, the Z signal is ``overwhelmed by QCD di-jet production'' that are induced by interactions, such as:
\begin{itemize}
    \item $gg$ $\rightarrow$ $gg$
    \item $qg$ $\rightarrow$ $qg$
    \item $q\bar{q} \rightarrow q\bar{q}$ (such as the t-channel gluon exchange interaction from Figure 1)
\end{itemize}

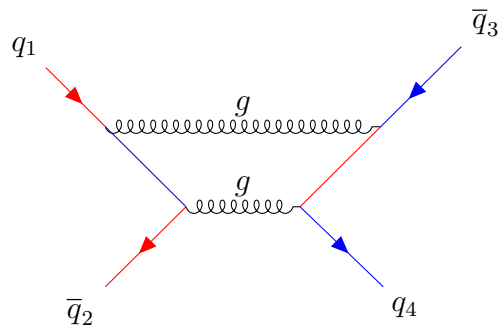
\begin{figure}[b!]
\centering
\begin{tikzpicture}
  \begin{feynman}
    \vertex (a) {\(q_{1}\)};
    \vertex [below right = of a ] (mid1);
    \vertex [right = of mid1] (mid2);
    \vertex [below right = of mid1] (b);
    \vertex [right = of b] (c);
    \vertex [above right = of c] (mid2);
    \vertex [above right = of mid2] (f1){\(\overline q_{3}\)};
    \vertex [below left = of b] (i2){\(\overline q_{2}\)};
    \vertex [below right = of c] (f2){\(q_{4}\)};

    \diagram* {
      (a) -- [fermion, red] (mid1) -- [gluon, edge label = $g$] (mid2),
      (mid1) -- [blue] (b) -- [gluon, edge label = $g$] (c) -- [red] (mid2) -- [anti fermion, blue] (f1),
      (b) -- [fermion, red] (i2),
      (c) -- [fermion, blue] (f2),
      
    };
  \end{feynman}
\end{tikzpicture}
\vspace{-2mm}
\caption{NLO Feynman diagram for the interaction $q\bar{q}$ $\rightarrow$ $q\bar{q}$. Gluon emission of a initial state quark leads to a colour singlet. This interaction is another source of interference.}
\label{fig: fig_2}
\end{figure}

At next-to-leading-order (NLO), the s-channel gluon QCD interaction has two more vertices, due to the emission and absorption of another gluon by the initial and final states respectively; this is visualised in Figure \ref{fig: fig_2}. In comparison to the LO s-channel QCD interaction, this diagram is a colour singlet, thus, this interaction also interferes with the s-channel electroweak process. Although this interaction is simulated in the MC, the interaction is simulated without considering interference with electroweak processes. In a recent paper \cite{Paper_7} from 2022, the role of colour octets producing colour singlets is explored, the paper states, ``the necessity of including the color-octet states in predictions of inclusive production at hadron colliders has been borne out in extensive experimental analyses''\cite{Paper_7}. Indeed, as highlighted, this role is an important factor which is not yet fully understood.

\begin{figure}[t!]
\centering
\includegraphics[width=150mm]{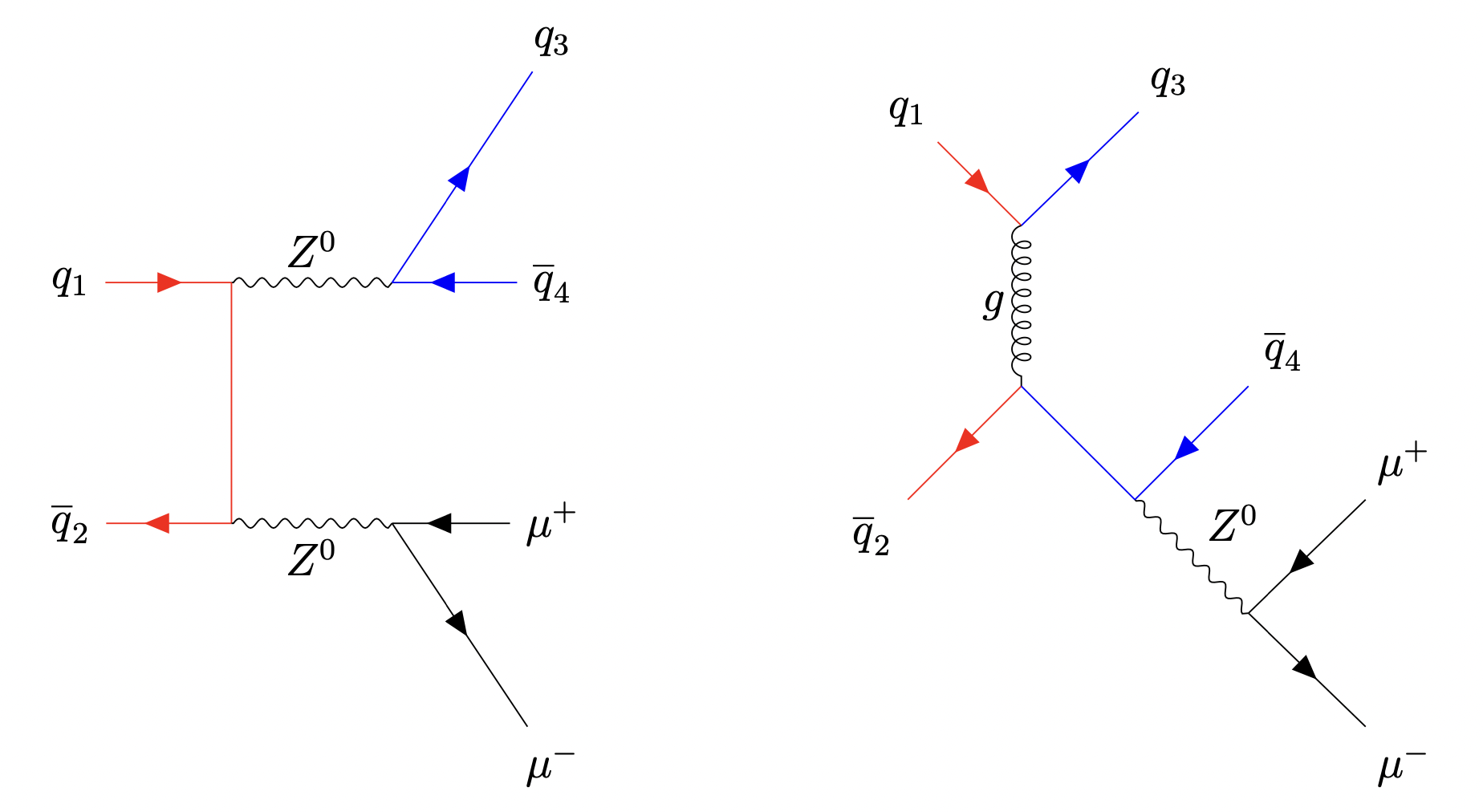}
\caption{a) LO signal diagram: $q\bar{q} \rightarrow$ ZZ $\rightarrow$ $q\bar{q} \mu\bar{\mu}$. b) LO QCD background diagram: $q\bar{q} \rightarrow$ Z$q\bar{q} \rightarrow$  $q\bar{q} \mu\bar{\mu}$, where the $q\bar{q}$ come from QCD processes. These two interactions interfere.}
\label{fig: fig_3}
\end{figure}

Now, a couple remarks on experimental techniques are made. The simplest experimental route to obtaining the signal is to study diboson pairs, ZZ. One of the Z bosons decays to a mu-mu pair, Z $\rightarrow$ $\mu\bar{\mu}$. This acts as trigger, initiating the recording of data. The other Z boson decays to the $q\bar{q}$ final state that is required; the full Feynman diagram is shown in Figure \ref{fig: fig_3}a. The use of triggers in this way have been previously used in papers such as Ref. \cite{Paper_8} to find evidence for Higgs boson decay to a quark-antiquark pair. The paper noted that the requirement of a trigger ``suppresses backgrounds with non-resonant lepton pairs, such as $t\bar{t}$ and multi-jet production'' \cite{Paper_8}. It is important to note here that electrons can also be used as a trigger in this way. However, muons are more precisely measured than electrons. This is because muons are the least interacting charged particles, hence they lose the least energy when interacting with the detector material. Electrons need to be treated more carefully, their energy needs to be recalculated due to photon emission when interacting with the detector. It is for this reason that muon triggers are favoured.

The experimentalist has the scope to distinguish final state particles, therefore this study will mainly consider $q\bar{q}$ $\rightarrow$ \emph{b}$\bar{b}\mu\bar{\mu}$ interactions. By demanding that jets are b-tagged, interactions with gluons in the final state are significantly suppressed. Furthermore, 20\% of Z decays are into a $b\bar{b}$ pair, whereas typical QCD jets have a 0.5\% chance of originating from a b quark. Both these factors allow for the enhancement of the Z signal by b-tagging.

\section{Goals}
The main goals of this project will now be stated:
\begin{enumerate}
  \item To experimentally observe Z boson production at low transverse momentum through their decay to b quarks.
  \item To compare the mass peak of the Z boson between MC and experimental data from the LHC. In particular, a statistically significant shift in the mass peak is hypothesised.

\end{enumerate}
These goals will be met through a series of sub-goals:

\begin{enumerate}
  \item To optimize the signal to background ratio through cuts and selections.
  \item To make the Z $\rightarrow$ \emph{b}$\bar{b}$ mass peak more distinct by recalculating the dib-jet mass using the recorded mass of the individual b-jets.
  \item To use function fitting to create a model that can extract an accurate signal from both MC and experimental data.
  \item To draw conclusions from statistical errors.
  
 Note: Although it is known that the jet systematics will have an effect on the signal mass peak, given the scope of the project, these effects will not be evaluated.
  
\end{enumerate}

\section{Methodology}
The methodology will now be presented. This section is split into several parts to allow for a coherent and logical structure, starting with details on the computational tools, MC samples and experimental data that are used. Focus is then shifted to the sub-goals which outline the methodology; each sub-goal is discussed in turn. Unless stated otherwise, all of the analysis in this section is done on the MC. The model is built using the MC with the vision of applying it to the experimental data to find a signal. At certain points, however, experimental data from the rejected sample is used for testing, but the experimental data that is used to form conclusions is not unblinded until the very end. This is to ensure there is no sub-conscious bias in any decisions taken.

\subsection{MC samples and experimental data}
Firstly, it is important to state here that the ROOT library is used to conduct all the analysis in this project. ROOT is a library developed by CERN and it runs on C++. 

The most significant MC samples to this project are the Z+jets, $t\bar{t}$ and diboson samples. Specific information, such as the degree to which these samples are corrected, is detailed in Table \ref{tab: tab_1}. Other MC samples are also used, as they are available, however it is important to note here that their contributions are very small, if not negligible. Thus, specific details about these samples are not too significant. In any case, these samples are corrected either to NLO or NNLO and are generated either by the POWHEG-BOX V2 or the SHERPA 2.2.1 package, refer to Table 1 from Ref. \cite{Paper_9} for full details. The physics behind these samples and details on the conditions used to generate them are explored in Appendix A. For the purposes of understanding the work presented, the most important takeaways are:
\begin{enumerate}
  \item The diboson sample describes the signal.
  \item The Z+jets is QCD background and accounts for both the biggest part of the background and the majority of total events.
  \item The $t\bar{t}$ also forms a significant part of the background, although is smaller than the Z+jets background.

\end{enumerate}

The experimental data used is from Run 2 of the ATLAS experiment, which ran from 2015 and 2018. The data was collected at a centre-of-mass energy of 13 TeV, with a luminosity of 139$fb^{-1}$. To reduce the data set, two pre-selected conditions were applied:

\begin{enumerate}
  \item Exactly two muons were demanded and,
  \item At least one jet.

\end{enumerate}

\begin{table*}[t]
\centering
\begin{tabular}{ c c c }
\hline 
 Sample & Generator & Cross-section order\\ 
\hline 
Z + jets & SHERPA 2.2.1 & NNLO \\ 
$t\bar{t}$ & POWHEG-BOX V2 & At least NNLO \\  
Diboson & SHERPA 2.2.1 & NLO \\  
\hline 
\end{tabular}
\caption{Refer to Ref. \cite{Paper_9}. Note: in addition to being corrected to NNLO, soft corrections have also been applied to the $t\bar{t}$ sample.}
\label{tab: tab_1}
\end{table*}

\subsection{Cuts and Selections}
 A large portion of the earlier stages of the project was dedicated to the first sub-goal: finding suitable cuts to increase the signal to background ratio. A whole range of potential variables to cut on were investigated. From the requirement of a trigger, two cuts were established early on:

\begin{enumerate}
  \item The mass of the dimuon pair has to be in agreement with the mass of the Z boson. That is, 81 GeV $<$ $m_{\mu \mu}$ $<$ 101 GeV. 
  \item The muons in the dimuon pair have to be of opposite charge.

\end{enumerate}

These conditions enrich the sample with events containing a dimuon pair arising from the leptonic decay of the Z boson. Next, in order to obtain a final state with a $b\bar{b}$ pair, a cut on the number of b-jets is desired.
Jets are split into two categories: central and forward. The difference between these two types of jets is based on the angle of detection of the jet. It suffices to know that all b-jets are central jets. Therefore, a cut of two or more central jets is desirable.  Upon inspection, it was observed that demanding exactly two central jets, as opposed to at least two, gives rise to a sharper signal peak. It was concluded that this gain outweighs the loss of a natural reduction in events. Therefore, a cut of exactly two central jets is demanded. It is noted here that the analysis of the dib-jet mass is done on data that has already been b-tagged, therefore it is given that these two central jets are b-jets.

Now, two options for cutting on the forward jets are considered. Either demand exactly 0 forward jets, or do not cut at all on the forward jets. The latter option allows for cases where initial or final state radiation has occurred, for example, when an incoming quark has given off a gluon before the main interaction occurs. This gluon will of course produce a jet through fragmentation. The former option is cleaner, in the sense that it demands there are only two jets in the final state, both of which are b-jets, as required. However, the former option also reduces the number of signal events by approximately 25\%. Therefore, the latter option is opted for to ensure there are as many signal events as possible.

\begin{figure}[b!]
\centering
\includegraphics[width=77mm]{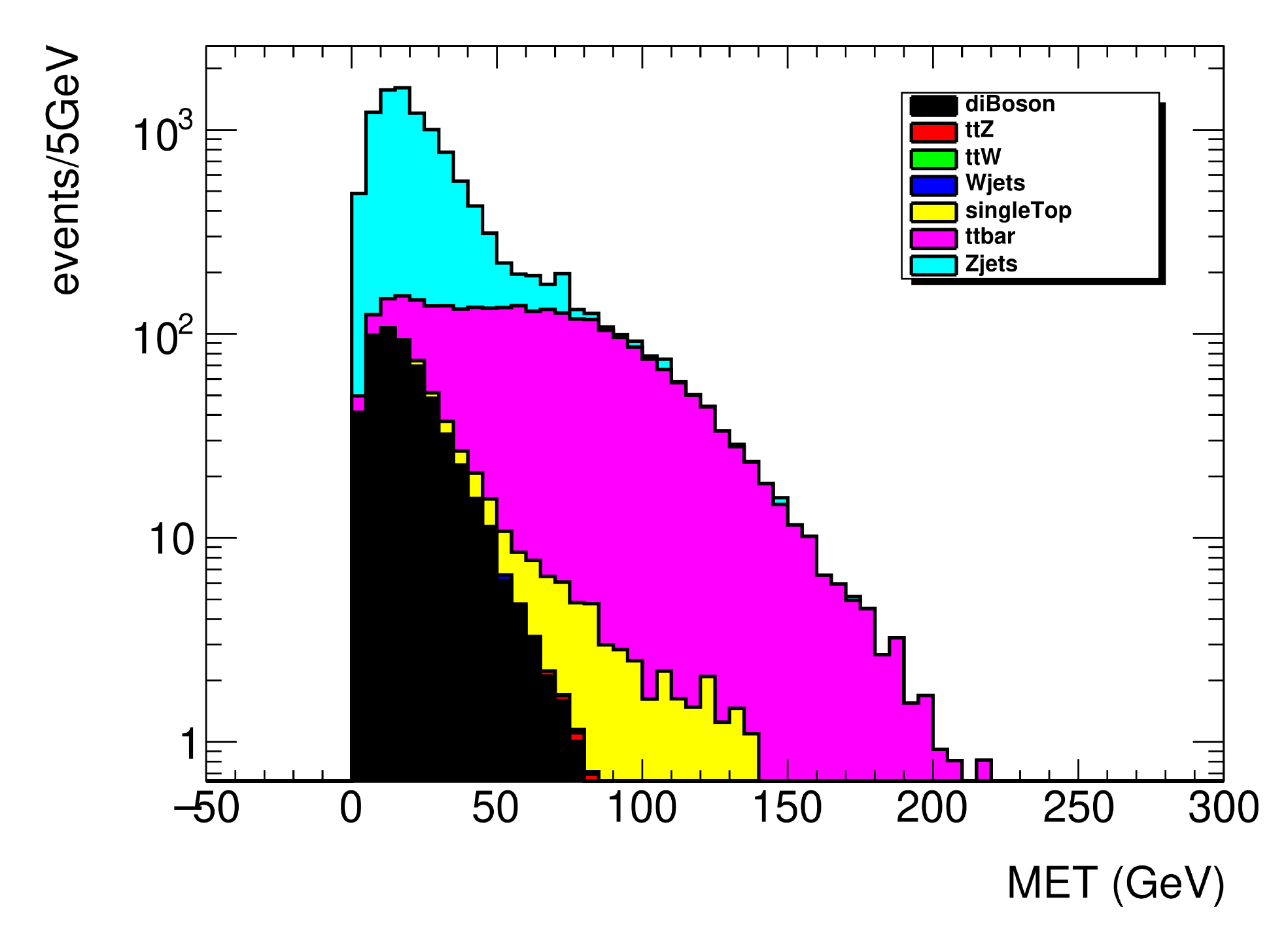} \hspace{3mm}
\caption{A graph of the MET. The black represents the signal, with background contributions in different colours, detailed by the legend. The condition that the dib-jet mass had to be between 70 and 100 GeV is demanded. This allows for the improvement of the signal to background ratio near the signal peak. As can be seen, the cut at 50 GeV discards a significant proportion of the $t\bar{t}$ background.} 
\label{fig: fig_4}
\end{figure}

Next, an algorithm was written to calculate the ratio of the number of signal events to the square root of the number of background events. This algorithm was used to find cuts by identifying maxima of this ratio. The only variable that gave a suitable cut was the missing transverse energy (MET). Upon inspection, it was observed that a cut at 50 GeV was optimal, as highlighted in Figure \ref{fig: fig_4}. More precisely, it is demanded that the MET is less than 50 GeV. Finally, since the aim is to study interference at low transverse momenta, a cut of $p_T < $ 100 GeV on the dimuon system is demanded.  A cut on the dimuon system gives a cut on the entire system; the reason the cut is made on the muon pair instead of the jets is simply because the properties of the muons are measured more precisely, as mentioned earlier.
100 GeV was chosen as the cut off point since this provides a fine balance between the number of events below and above this point, allowing for the rejected sample with $p_T > $ 100 GeV to be used as a test region for the model, albeit with worse performance than in the target region.

\subsection{Corrections to the dijet mass}
This section concentrates on the second sub-goal: recalculating the dib-jet mass to make the signal peak more distinct. In turn, this would make it easier for a signal to be detected in experimental data. At first, an attempt was made to reconstruct the mass of the b-jets by correcting for the MET using trigonometry. On inspection, the correction actually seemed to make matters worse: it caused the signal to background ratio to decrease and made the signal peak wider. This may be down to the nature of the kinematics at low $p_T$: as $p_T$ $\rightarrow$ 0, the b-jets will be produced back-to-back, probing only one direction. As a result, it is unclear where any MET should be assigned. This is one of the challenges that arises when low $p_T$ is considered: a technique which is common practice in higher $p_T$ regions fails to work. As a result, another potential solution was explored: the dib-jet mass was recalculated using the recorded mass of the individual b-jets. The following correction was tested:

\begin{eqnarray}
corrected\;dib\;mass = measured\;dib\;mass - \alpha (bjet1\;mass) - \beta(bjet2\;mass) + \gamma\,.
\label{equation8}
\end{eqnarray}
where $\alpha$, $\beta$ $\gamma$ are constants to be found such that the mean remains unchanged and the best results are obtained against the following criteria:

\begin{enumerate}
  \item  The percentage increase in the height of the signal peak is maximised.
  \item The percentage increase in the height of the Z+jets and $t\bar{t}$ is minimised.

\end{enumerate}
Intrinsically, there is a correlation between these two criterion: as the height of the signal peak increases, generally the height of the background increases too. A judgement has to be made regarding this trade-off. Nevertheless, the constants were varied to obtain a suitable combination of constants with respect to the criteria, achieving a 24\% increase in the height of the signal, a 11.5\% increase in the height of the Z+jets and a 20\% increase in the height of the $t\bar{t}$ sample. The increase of 24\% in the height of the signal is a significant improvement, but the 20\% increase in the height of the $t\bar{t}$ sample is undesirable.

\begin{figure}[b!]
\centering
\includegraphics[width=77mm]{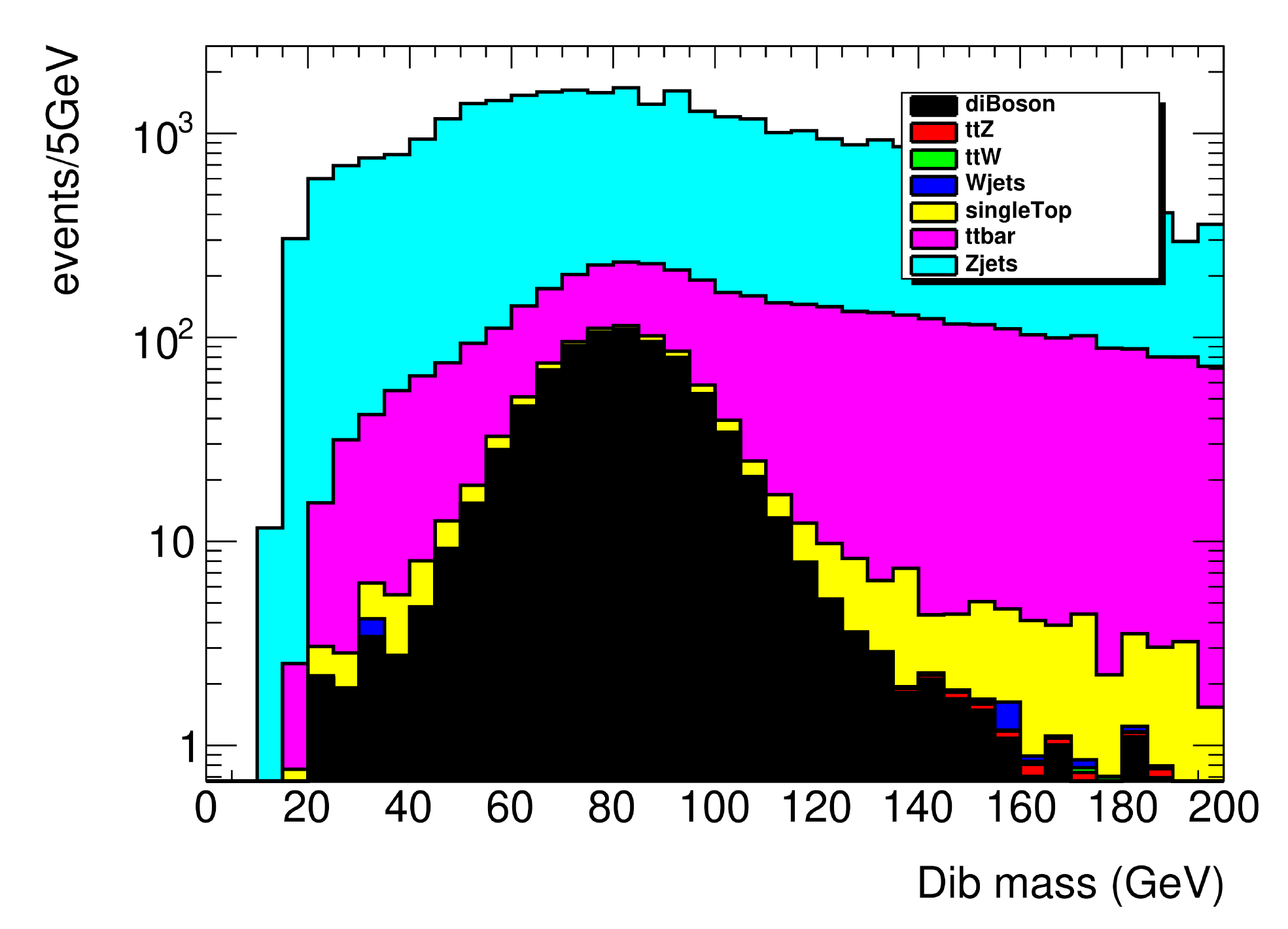} \hspace{3mm}
\includegraphics[width=77mm]{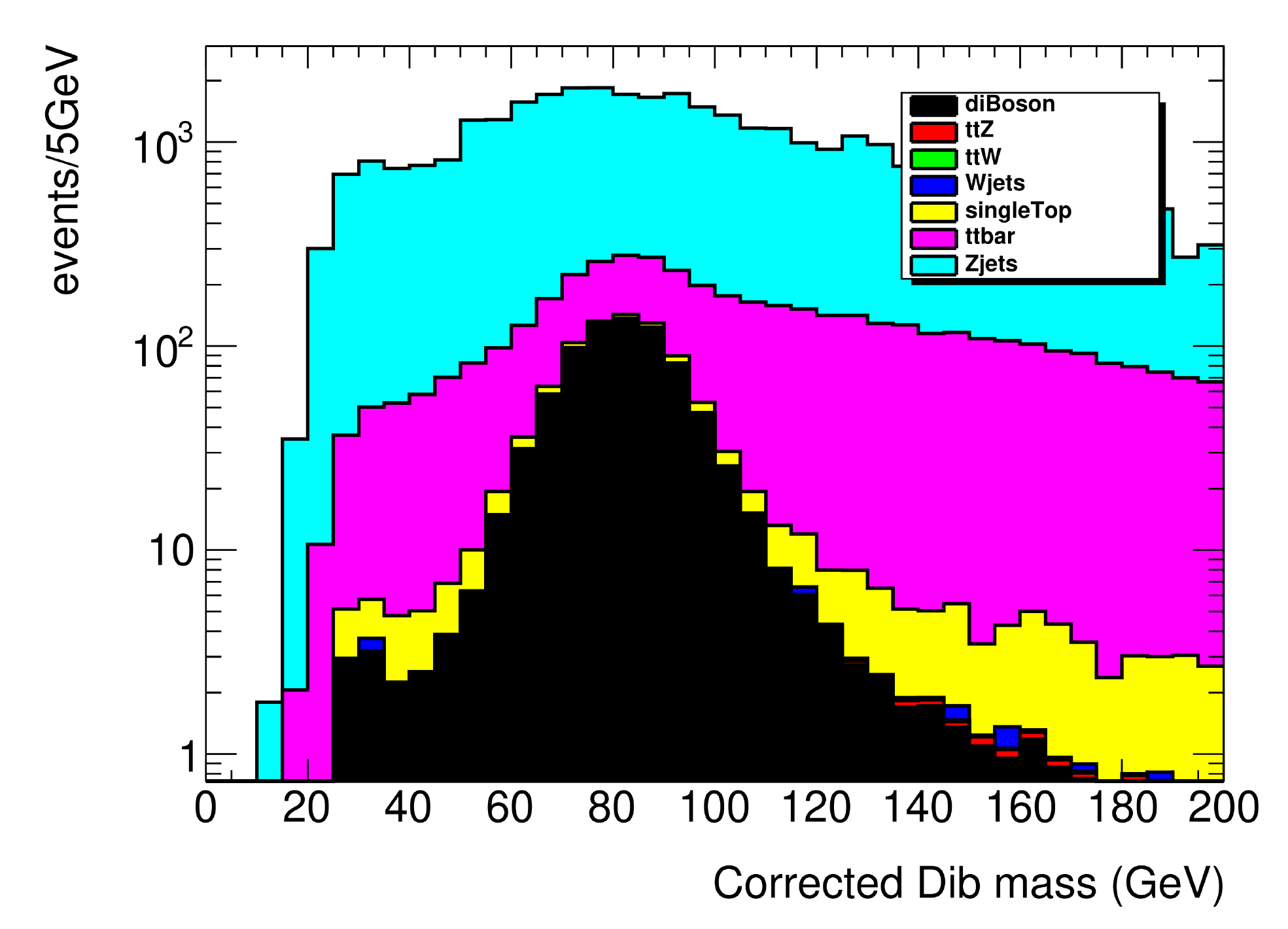}
\caption{a) Recorded dib-jet mass (left). b) Corrected dib-jet mass (right). The height of the signal peak increases and the width decreases after the correction is applied.}
\label{fig: fig_5}
\end{figure}

It was hypothesised that this was caused by heavy jets receiving large correction factors. As a result, the correction was only applied to jets with masses lower than 10 GeV. This cut off point was varied and 10 GeV was deemed as the most suitable. The mass of the b quark is approximately 4.5 GeV \cite{Paper_3}; jets with masses over 10 GeV are more than twice as massive as the mass of the b-quark, thus, from a theoretical perspective, it is reasonable to regard these jets as heavy jets.
This correction yielded an increase in the signal peak by 21\%, which is slightly lower than the 24\% achieved in the previous correction, but, the increase in the height of the Z+jets decreased from 11.5\% to 8.5\%. More significantly, the increase in the height of the $t\bar{t}$ sample decreased from 20\% to 6.6\% when compared to previous corrections where larger jets were also being corrected for. Therefore, this correction yields a significant increase in the height of the signal peak, while minimising increases in the background peak. Refer to Figure \ref{fig: fig_5} for a visual comparison of the dib-jet mass before and after this correction is applied. Any mention of the dib-jet mass from now will implicitly imply that this correction has been applied. The untagged Z+jets will be introduced and used later on; these jets will also be corrected using the same formula to ensure consistency.

\begin{figure}[t!]
\centering
\includegraphics[width=77mm]{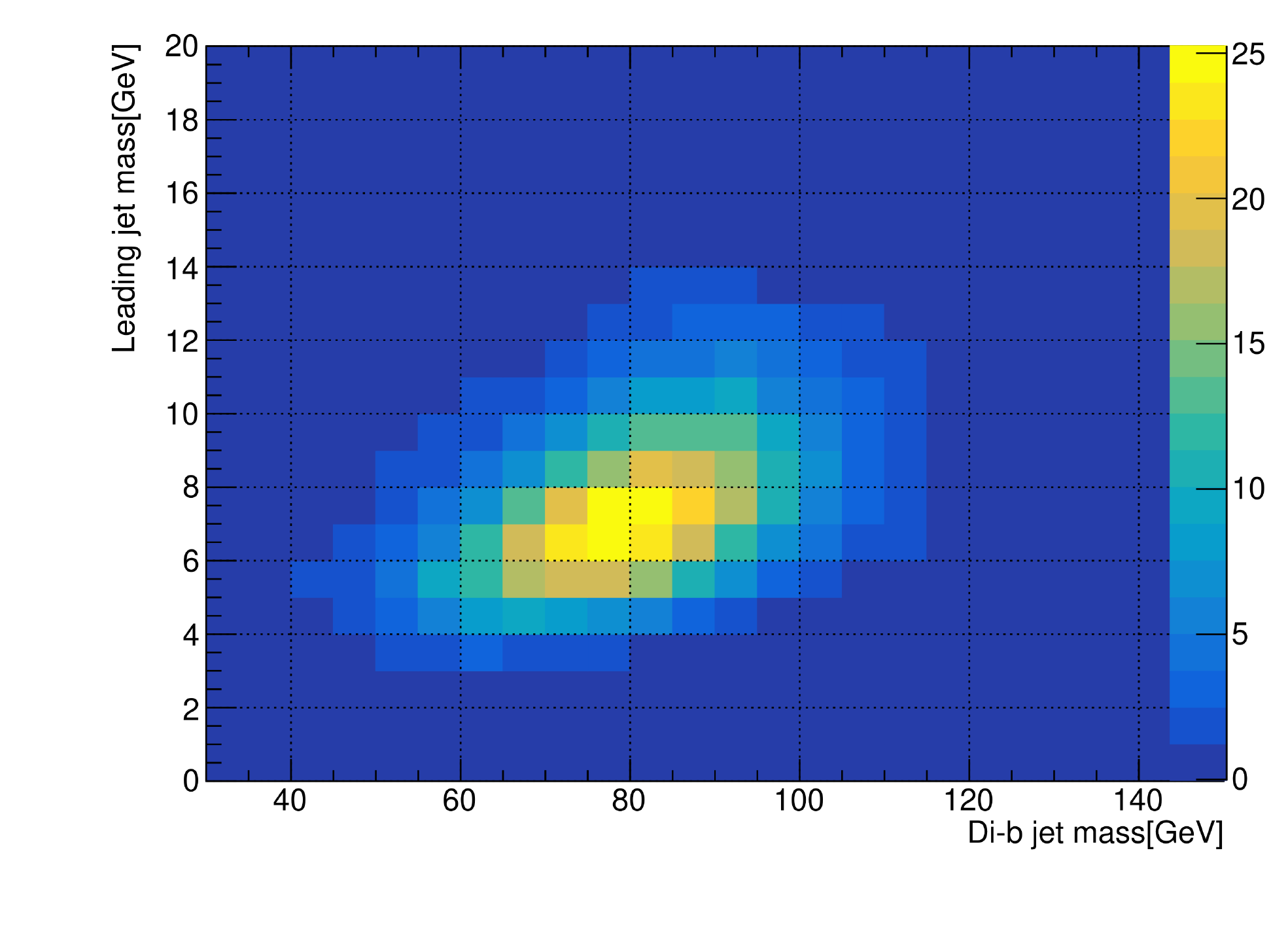} \hspace{3mm}
\caption{The 2D plot highlights the correlation between the dib-jet mass and the mass of the b-jet with higher $p_T$: the leading jet. This correlation explains why the mass of the b-jets can be used to correct the dib-jet mass to give a sharper signal peak.}
\label{fig: fig_6}
\end{figure}

Two natural questions arise from this sort of correction. Firstly, why does the dib-jet mass need to be corrected? The b quark can decay into a charm quark and a $W^{-}$ boson, neutrinos can then be produced through further decay. These neutrinos are not registered by the detector and hence carry energy away, reducing the recorded energy and mass of the jets. Conversely, extra pile-up energy from another jet can lead to a increase in the mass of the dib-jet system. These corrections attempt to compensate for these scenarios. The second question is, why does a correction based on the recorded mass of the individual b-jets work? This is due to the fact that the individual jet $p_{T}$ is correlated with the mass of the jet and the di-jet pair mass is correlated with the $p_{T}$ of the individual jets. Thus, the dib-jet mass is correlated to the mass of the individual b-jets, as can be seen in the 2D plot in Figure \ref{fig: fig_6}.

\subsection{Functional form of the signal}
Next, attention is focused on the third sub-goal. The final model that is built is a combination of three sub-models, namely: a functional form of the signal, a model for the dib-jet mass of the Z+jets background and a model for the sum of the other background contributions, with the latter two making up the background model. It is this final model which will be fitted to the dib-jet mass distribution found in experimental data.
Some terminology is introduced for clarification and simplification. The sum of all MC samples will be called \emph{fake data}. Essentially this collection simulates the experimental data. The experimental data will simply be referred to as \emph{data} from henceforth.

\begin{figure}[t!]
\centering
\includegraphics[width=77mm]{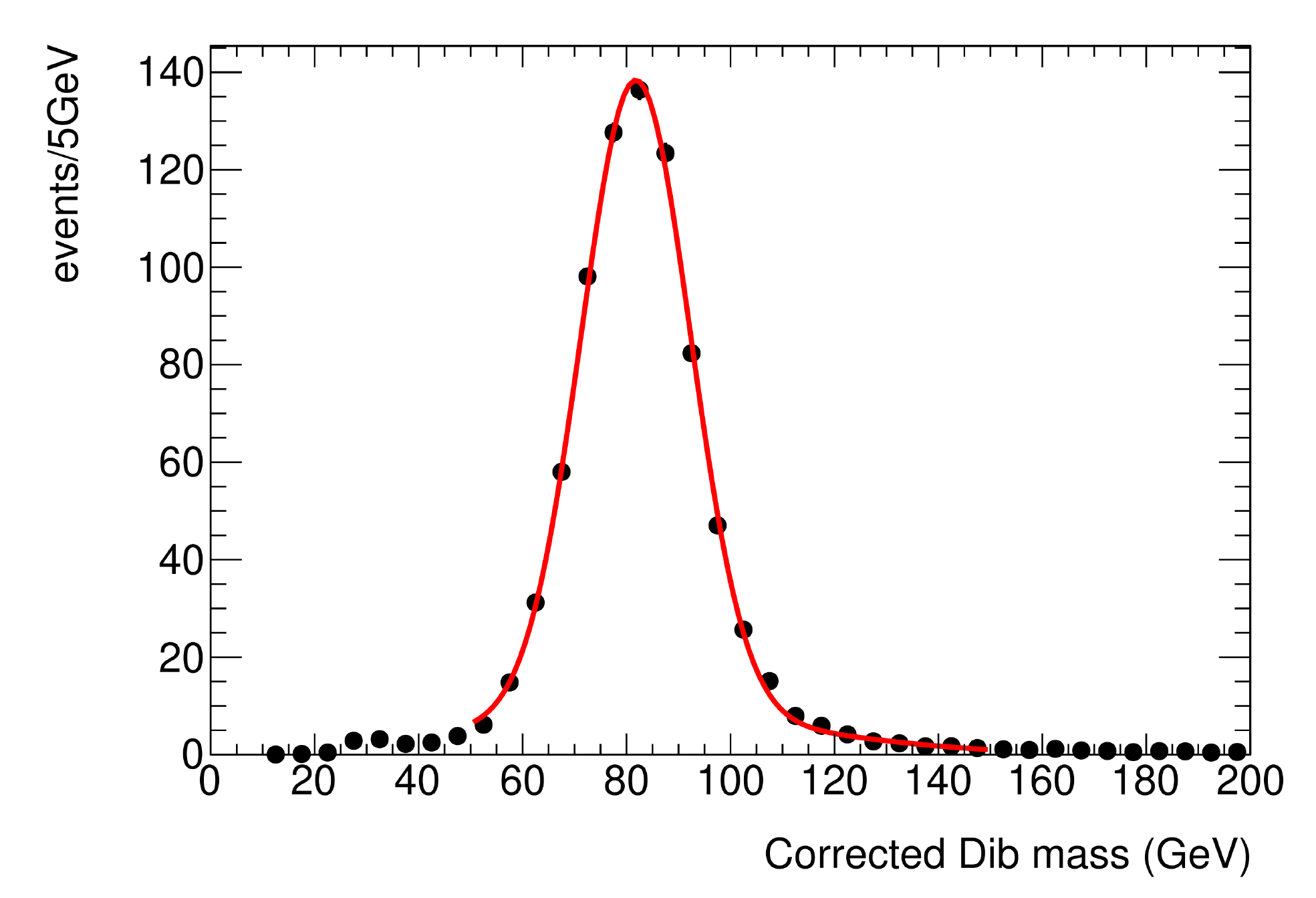} \hspace{3mm}
\caption{The sum of two Gaussian distributions with equal mean is fitted to the dib-jet mass of the diboson sample. As can be seen visually, the fit models the signal peak in the MC well. It is noted that the range of the fit is 50-150 GeV, this will be discussed in the next subsection.}
\label{fig: fig_7}
\end{figure}

Now, the first step in the process is to use the diboson MC sample to find a suitable function that can model the signal. Different functional forms were fitted to the dib-jet mass of the diboson MC and compared to each other, both visually and based on the returned chi-squared ($\chi^{2}$) for each fit. The $\chi^{2}$ allows for a comparison for the goodness-of-fit between different fits. A good fit is achieved using a linear combination of two Gaussian distributions with equal mean. This functional form has five parameters, one of which is the mean. Practically and theoretically, this mean represents the mass peak of the signal. Therefore, the mean parameter will be freed in the final fit and the returned best fit value will give the value of the mass peak in the data. Through a comparison with the mass peak in the diboson MC, a shift in the mass peak can then be assessed, with careful consideration of the errors associated with the returned best fit values. The other four parameters will be fixed to the values obtained when the function is fitted to the dib-jet mass of the diboson MC. These can be thought of as control variables, the shift in the mean is the variable being tested for. A visualisation of the fit can be seen in Figure \ref{fig: fig_7}. The returned best fit value for the mean is approximately 81.8 $\pm$ 0.08 GeV. This is about 10 GeV lower than the mass of the Z boson; this reduction in mass is due to the decay into neutrinos, as discussed previously.

\subsection{Modelling the background}
The background model comprises two parts: the Z+jets and the sum of the other background contributions. The latter is discussed first. The most significant contribution to this sample is the $t\bar{t}$ MC. The sum of the remaining background MC samples, excluding $t\bar{t}$, make up a very small proportion of the overall number of events. Their presence is statistically insignificant. 
Furthermore, the reliability of these samples and of the $t\bar{t}$ sample, is not being challenged. As such, these MC samples are directly used in the model to account for this background.

\begin{figure}[t!]
\centering
\includegraphics[width=77mm]{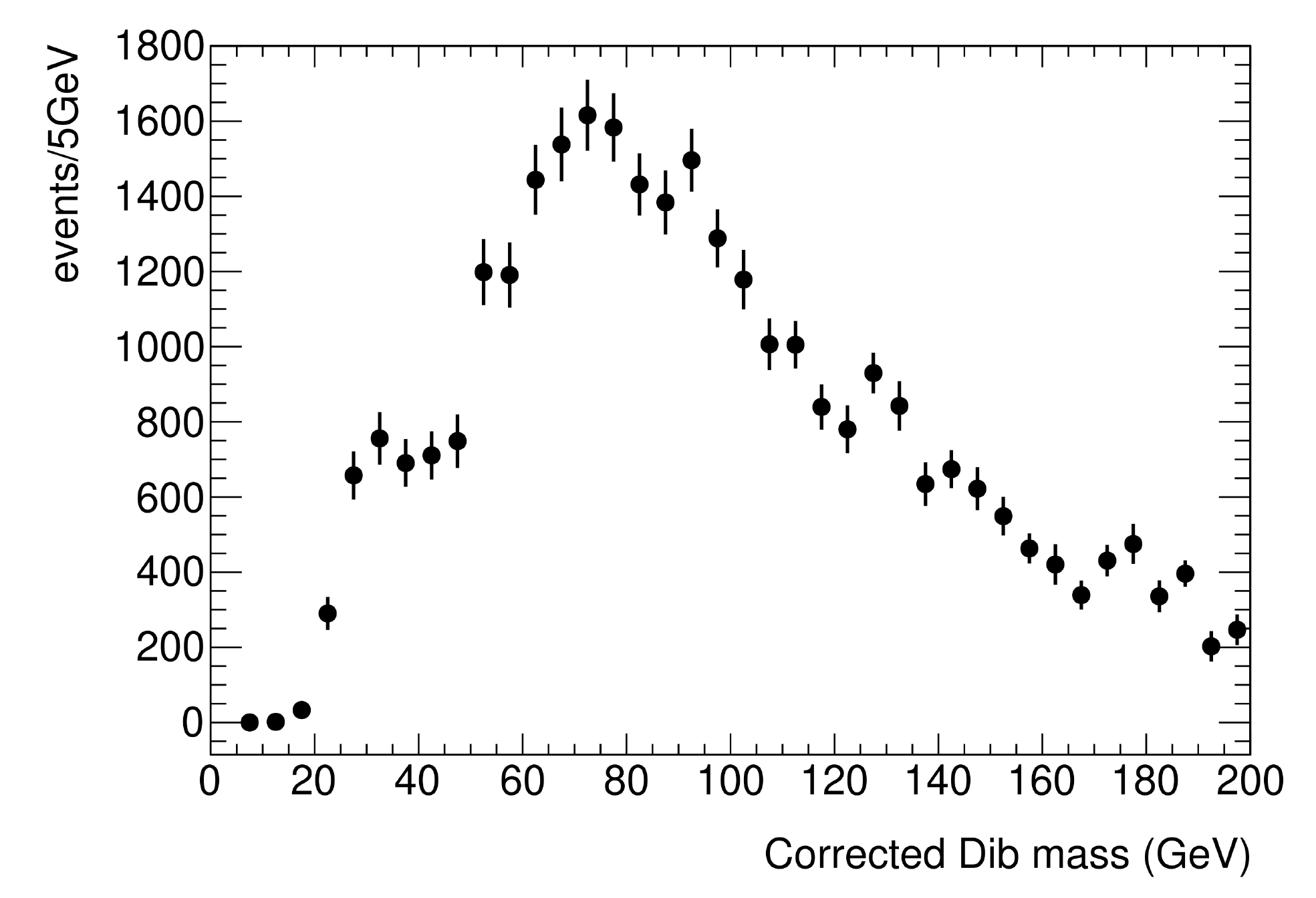} \hspace{3mm}
\includegraphics[width=77mm]{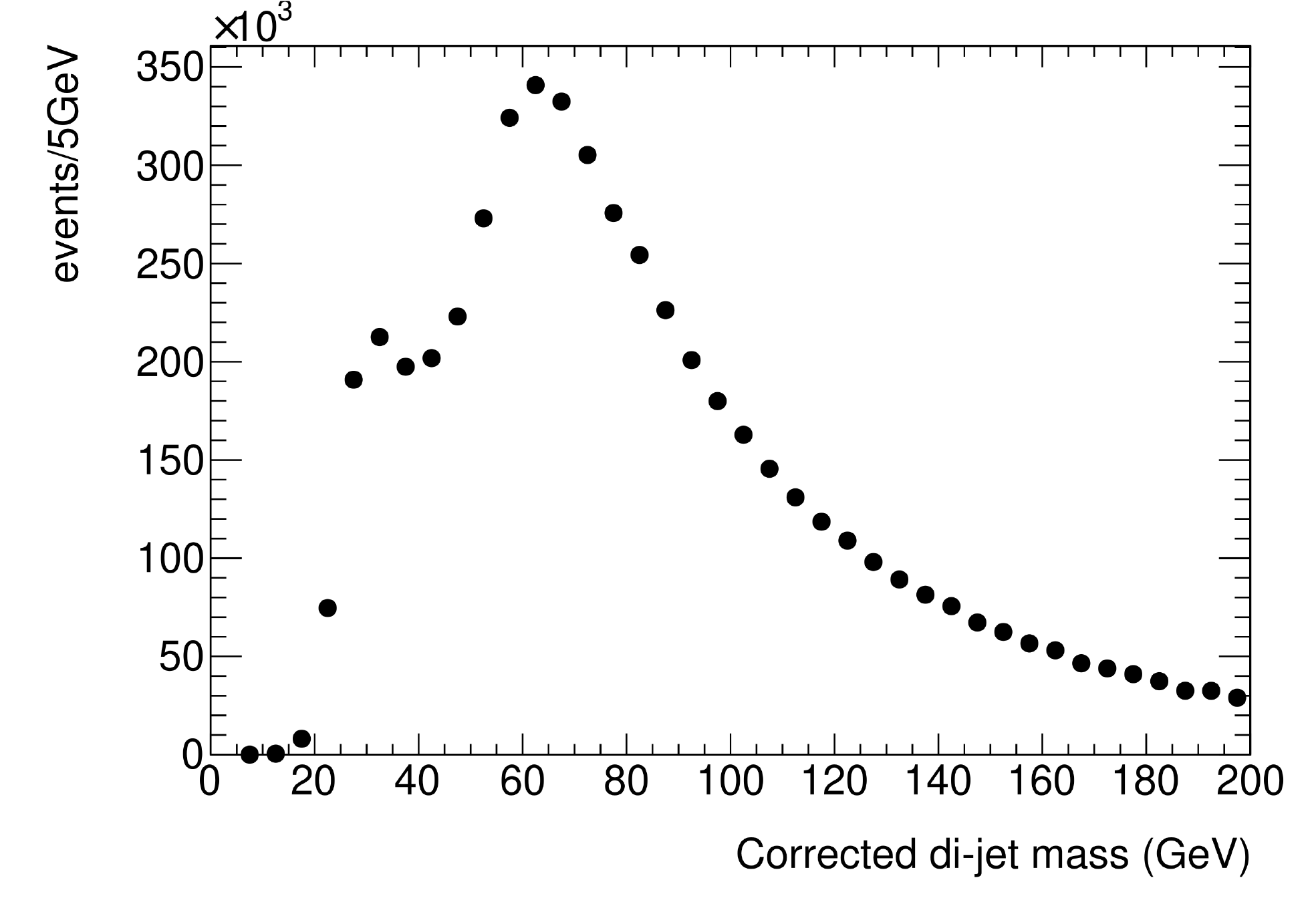}
\caption{a) Dib-mass of Z+jets sample (left). Large errors are observed. Random fluctuations in the distribution are highly likely due to a lack of statistics rather than genuine physics. b) Untagged di-jet mass of Z+jets (right). Note that 0 b-tags are demanded for this sample so that events are not being used twice. The shape of this distribution is smoother, although both distributions have similar shapes and characteristics. For instance, they both show a small peak at low masses, this may be due to the minimum jet $p_{T}$ threshold of 20 GeV. To make modelling easier, the lower bound for the fit will be moved up to 50 GeV.}
\label{fig: fig_8}
\end{figure}

Now, modelling of the dib-jet mass of the Z+jets is explored. Having looked at the Z+jets MC, it was observed that the error on the Z+jets MC is approximately five times greater than the square root of the number of events, which is equal to the error on data. The error on the Z+jets MC is a known problem and can make the modelling of the Z+jets particularly difficult. 

In an attempt to diminish the effects of these errors, the following method is pursued. A scaled-down version of the untagged di-jet mass of the Z+jets is used to model the b-tagged Z+jets. This is done through finding a suitable function to fit to the following ratio distribution: the dib-jet mass distribution of the Z+jets divided by the untagged di-jet mass distribution of the Z+jets. Then, this \emph{ratio function} is used to scale down the untagged Z+jets. The untagged Z+jets sample has more events than the tagged, thus has a lower error. It can be used in this way since it has a very similar distribution to the tagged Z+jets, this is because the underlying physics is also very similar in these two cases. Plots of the tagged and untagged di-jet Z+jets distributions are shown in Figure \ref{fig: fig_8}.

When the final fit is done on data, the scaled untagged Z+jets from data is used to represent the tagged Z+jets data. The data is not split into different contributions, therefore to obtain the untagged Z+jets in data, the untagged diboson sample and the untagged $t\bar{t}$ sample are taken away from the untagged data, essentially leaving the Z+jets. The perfectionist would argue that the other background MC contributions should also be subtracted, but as argued before, their contribution is negligible. It is noted that in this method, the tagged Z+jets is not represented by a functional form based on the tagged Z+jets MC, therefore this method relies much less heavily on the statistics of the tagged Z+jets MC sample.

\begin{figure}[t!]
\centering
\includegraphics[width=77mm]{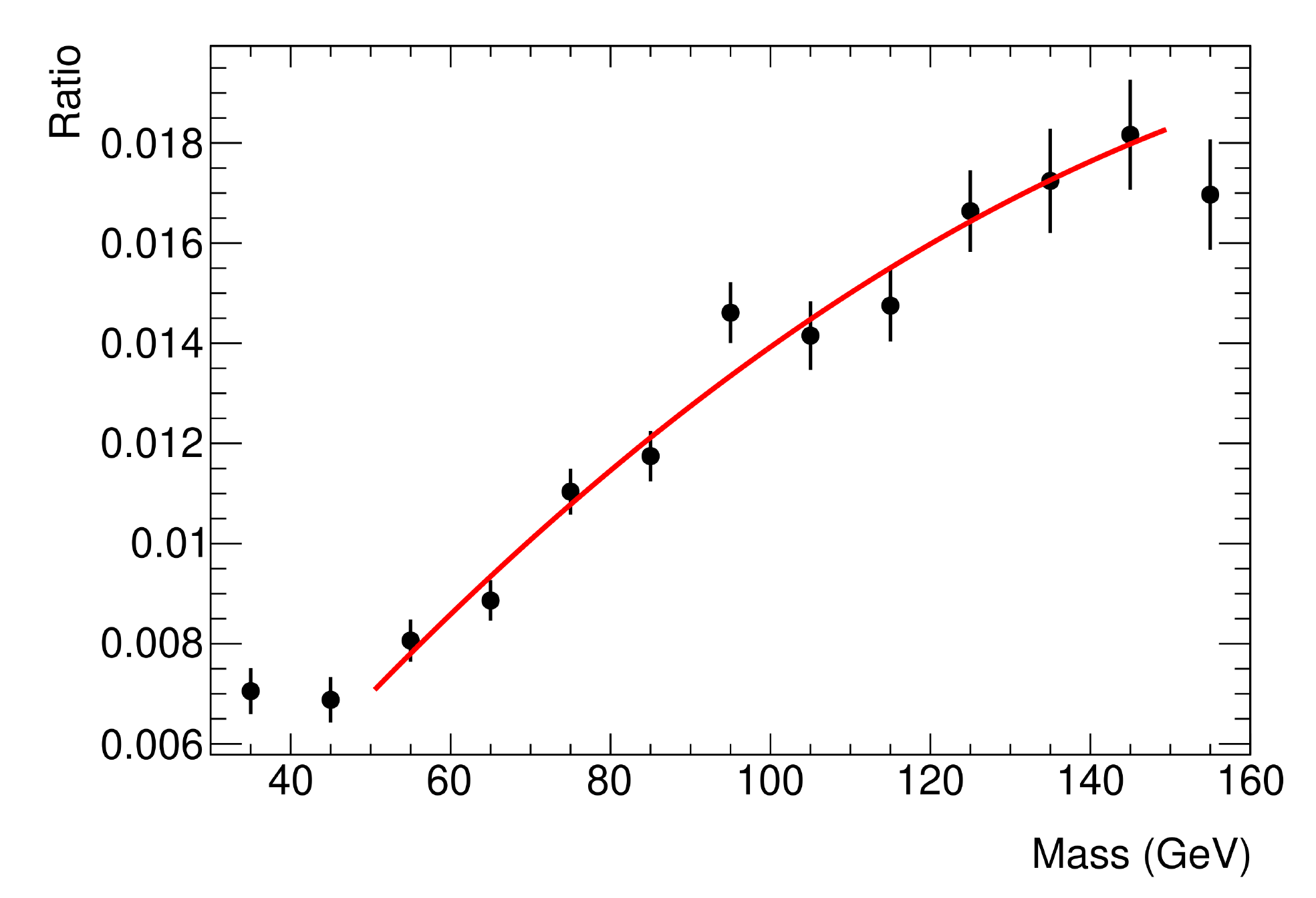} \hspace{3mm}
\caption{A second order polynomial is fitted to the ratio distribution within the range of 50-150 GeV. The fitted function is shown in red.}
\label{fig: fig_9}
\end{figure}

Now, the functional form of this ratio is presented. Polynomials of different orders were fitted to this ratio and, again, compared for goodness-of-fit. The problem here, however, is more delicate than it was for finding a functional form for the signal. By adding more parameters, the ratio is modelled better with a lower $\chi^{2}$. However, one does not want to add so many parameters such that the background model can describe every detail in the data, essentially making it difficult to find a signal. A fine balance must be struck. In addition, upon inspection of this ratio function, between the range of 160-200 GeV the distribution appeared to fluctuate, with rather large errors. This required higher order polynomials to describe. As these masses are significantly above the expected mass peak of the Z boson, the range of the fit can be reduced. As a result, to account for all these factors, a second order polynomial is used to model the ratio function with an upper bound of 150 GeV.
The ratio distribution and its functional form can be observed in Figure \ref{fig: fig_9}.

Although the method of finding a functional form of the tagged Z+jets MC has been rejected in this study, this is not to say that this method does not also have its benefits. For instance, this method allows for the untagged data to be used as a test region for the final model. In stark contrast, this is not possible in the method that is being pursued, since the untagged data will be used in the final fit. Therefore there are positives and negatives to each method and one needs to weigh them against each other to come to a reasoned conclusion. It is also noted that the method being pursued, using the ratio distribution, is inspired by a recent paper \cite{Paper_10}, in which a similar method is adopted.

\subsection{Testing the model}
The signal and background models can now be combined to form the final model. Now, several tests can be run on the model to evaluate its performance and to make any changes, if necessary, before the data is unfolded.

\begin{figure}[t!]
\centering
\includegraphics[width=77mm]{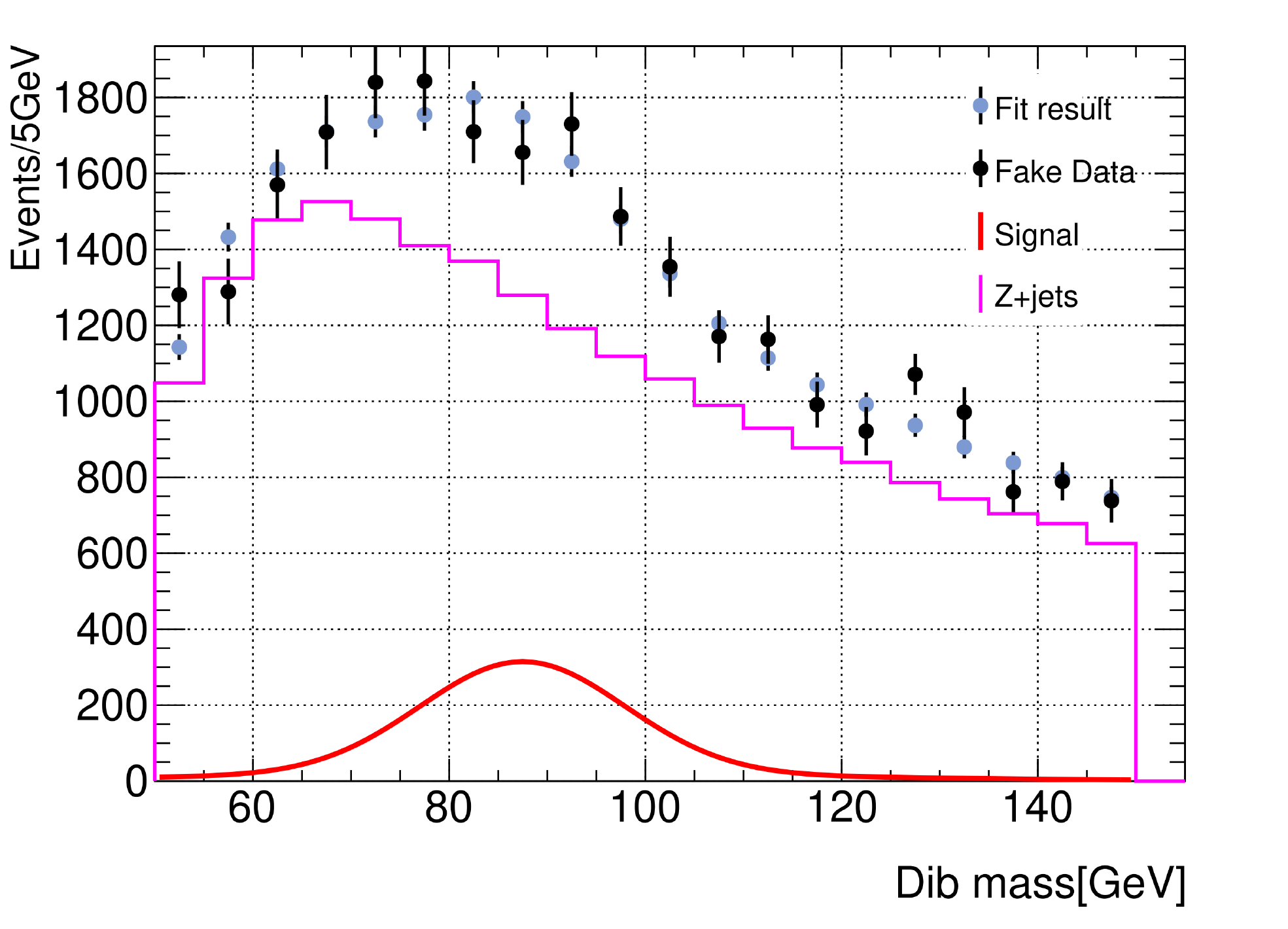} \hspace{3mm}
\caption{Visualisation of the fit to the dib-jet mass fake data. The fit is returning approximately twice as much signal as expected. }
\label{fig: fig_10}
\end{figure}

\subsubsection{Test 1: Apply model to dib-jet mass MC}
The model was tested on the dib-jet mass fake data. A visualisation of the fit is given in Figure \ref{fig: fig_10}. The parameter, \emph{musignal} represents how much signal is detected. As the functional form of the signal  was based on the diboson MC, the best fit value for musignal should be compatible with 1 when fitted to the fake data. However, a musignal of 2.3 $\pm$ 0.3 is returned. Therefore the fit is returning a fake signal, however, this is likely due to the error on the Z+jets MC. The fitted mean is also higher than expected, at a value of 87.5 $\pm$ 1.2 GeV. When the functional form of the signal was fitted to the diboson MC, a mean of 81.8 $\pm$ 0.08 GeV was returned, therefore this increase in the mean can again be due to large errors in the fake data. When comparing the mean from data to the MC mean, the mean fitted to the diboson MC sample will be used, as opposed to the fitted mean to fake data.

Overall, from Figure \ref{fig: fig_10} it is clear that the fit does describe the fake data reasonably well. However, applying the model to fake data did not provide any useful information about its performance in regards to detecting an accurate level of signal, or returning a reasonable value for the signal mass peak. The errors in the Z+jets MC are too large to provide a useful test.

\subsubsection{Test 2: Linearity test to fake data}
Although the fitted musignal was returning a fake signal, the sensitivity of the model to the injection of more signal can still be tested. The diboson MC in the fake data was scaled by a factor of 2. If the model is working well, the increase in the fitted musignal should be compatible with 1. Indeed, this is the case: the musignal increases from 2.1 $\pm$ 0.3 to 3.1 $\pm$ 0.3, which is compatible with an increase of 1 with a 68\% confidence level.

\subsubsection{Test 3: 1 b-tagged region}
In this project, the main focus of the study is on the dib-jet region. No data from the 1 b-tagged region will be used to draw conclusions. Thus, this region can be used as a test region for the model. Why does this sample have a potential to be a good test region? Since Z decay into a $b\bar{b}$ pair is being studied, if one jet is b-tagged, theoretically there should be another b-jet. However, the accuracy of b-tagging is approximately 77\%, therefore a small proportion of jets originating from b quarks will not be picked up by the tagging process. This works to reduce the number of events in the  dib-tagged region compared to the 1 b-tagged region. Therefore this region could be a good indicator as to how well the model is performing. On the other hand, there may be other reasons contributing to the fact that only one jet is b-tagged. For instance, consider an event with one gluon jet and two b-jets. The two b-jets may be so close together that they are registered as one jet, or, alternatively,  the energy of one of the b-jets may be so low that it is not even registered as a jet. The physics of these events is intrinsically different as there is a gluon jet in the final state arising from the interaction. Indeed, single b-tagged events like these cannot be simply explained by the imperfection of the b-tagging process, and, furthermore, these events will change the distribution of the 1-tagged region, potentially making it an invalid test region. Nevertheless, this region was tested to see if anything could be learnt.

Since the signal model found earlier was unique to the dib-jet region, in order to test the model fairly, a suitable signal model was found for the 1 b-tagged region. One can think of this as a test on the whole methodology pursued in this project, as opposed to a test solely on the model built. When the final fit was run on 1 b-tagged fake data, a musignal of 3 $\pm$ 0.5 was fitted. However a musignal of 1 was expected since the functional form of the signal was based on the diboson MC, as before. An optimist would argue that this could be down to the statistics on the Z+jets MC. Therefore, a fit to data was then carried out. This can be done as the 1 b-tagged data will not be used in the final fit or to draw any conclusions. When the fit to data was carried out, a musignal of 3.6 $\pm$  0.6 was returned. Therefore, the fit to data returns a value for musignal which is compatible to the value returned when fake data is fitted. On one hand, this is a success.
However, this result also suggests that the large value of musignal returned is not simply due to the statistics on the Z+jets MC sample. Hence, this region is not a good test for the model.

\subsubsection{Test 4: $p_{T}$ $>$ 100 GeV region}
One final test on the $p_{T}$ $>$ 100 GeV sample was done in an attempt to carry out a successful closure test. As before, the methodology was carried out and a fit to both fake data and data was executed. The number of events decreased by approximately 80\%, compared to the $p_{T}$ $<$ 100 GeV region, with the number of signal events dropping by the same amount, so errors were expected to be bigger. The fit to fake data fitted a value of 2.3 to musignal, with an error of 0.5. This error is significantly larger than its counterpart in the $p_{T}$ $<$ 100 GeV sample.

\begin{figure}[b!]
\centering
\includegraphics[width=77mm]{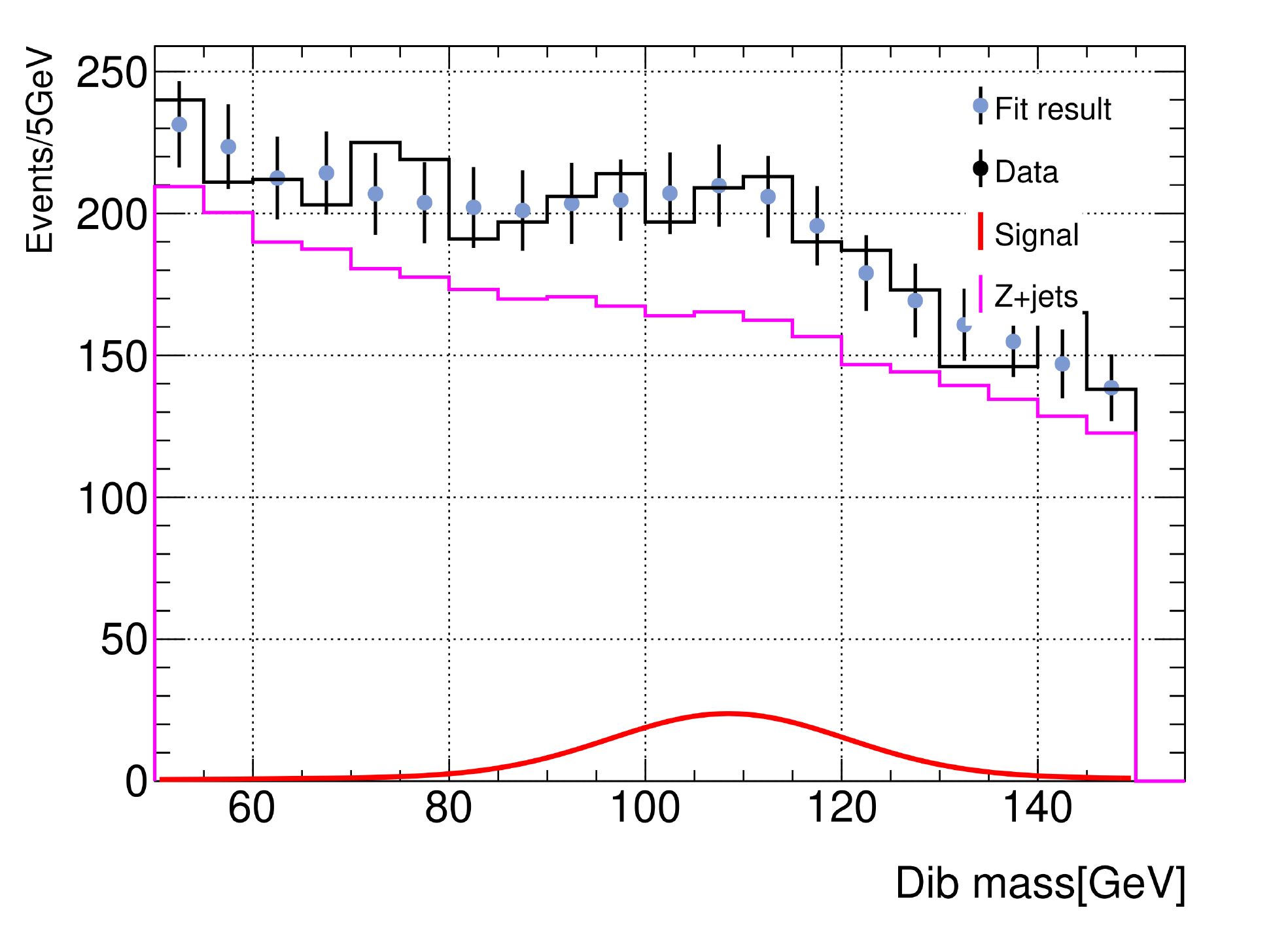} \hspace{3mm}
\caption{Fit to data in the $p_{T}$ $>$ 100 GeV region. The mass peak is shifted signficantly to the higher masses. This is potentially due to a Higgs background.}
\label{fig: fig_11}
\end{figure}

A fit to data gave an unforeseen result, a visualisation of which is given in Figure \ref{fig: fig_11}. The fitted mean was 108 $\pm$ 5.5 GeV, significantly larger than the mass of the Z boson. This could be due to considerable contributions from the decay of the Higgs boson in this  $p_{T}$ region. The mass of the Higgs is approximately 125 Gev \cite{Paper_10} and may explain the shift in the signal peak.
For this reason, the $p_{T}$ $>$ 100 GeV region is not an appropriate test space.
It is also noted that Higgs decay is not simulated in the diboson MC, therefore this shift is not seen in the fake data.
The contributions from Higgs decay is significantly suppressed in regions of lower $p_{T}$.

Given the time limitations of the project, as many tests were done as possible, yet no test seem to offer conclusive backing that the model will work sufficiently well on the dib-tagged data. Nevertheless, the fit to data was finally done and the results are presented in the following section.

\section{Results}
Before the results are presented, a remark is made. The NLO diboson MC samples have an inclusive cross-section of 14$pb$, whereas the data is closer to the more precise NNLO predictions, at a cross-section of 17$pb$. Therefore, it is expected that there is around 20\% more signal in the data compared to the fake data. Although interference is expected to shift the mass peak, it should not alter the normalisation. Therefore, when data is fitted, if a value of musignal compatible with 1.2 is fitted, then this would be a good indication that the fit is performing well.

Now, the results will be presented. Three fits to data were executed:

\begin{enumerate}
  \item  A fit with musignal and the mean freed,
  \item A fit with musignal fixed at 1.2 and the mean freed,
  \item A fit with musignal and the mean fixed at 1.2 and 82 GeV respectively.
\end{enumerate}

The first fit has the least constraints and therefore is the best test for interference in the data. The second fit attempts to help the fit by fixing the normalisation to what is expected. The third fit gives an insight to the goodness-of-fit when the most help is given.

\subsection{First Fit}
When fitted to data, a value of 1.4 $\pm$  0.4 is fitted to musignal; this is compatible with the expected value of 1.2. None of the tests done on the fit in earlier sections gave any conclusive evidence on the ability of the fit to detect the signal with enough sensitivity. However, this is a clear positive triumph for the fit. The mean was fitted with a value of 82.7 $\pm$  2.1 GeV.

\begin{table*}[b!]
\centering
\begin{tabular}{ c c c}
\hline 
 Parameter & Fitted value to MC & Fitted value to data\\ 
\hline 
Musignal & 2.1 $\pm$ 0.3 & 1.2 (fixed) \\ 
Mean (GeV) & 81.8 $\pm$ 0.08 & 82.7 $\pm$ 2.3 \\  
\hline 
\end{tabular}
\caption{Results from the second fit. Musignal is fixed at 1.2 for data. The means are compatible with each other, thus there is no evidence for interference. Note that the fitted mean value to MC is the mean value obtained when the functional form of the signal is fitted to the diboson MC, not from the overall fit to fake data.}
\label{tab: tab_2}
\end{table*}

\subsection{Second fit}
\begin{figure}[b!]
\centering
\includegraphics[width=77mm]{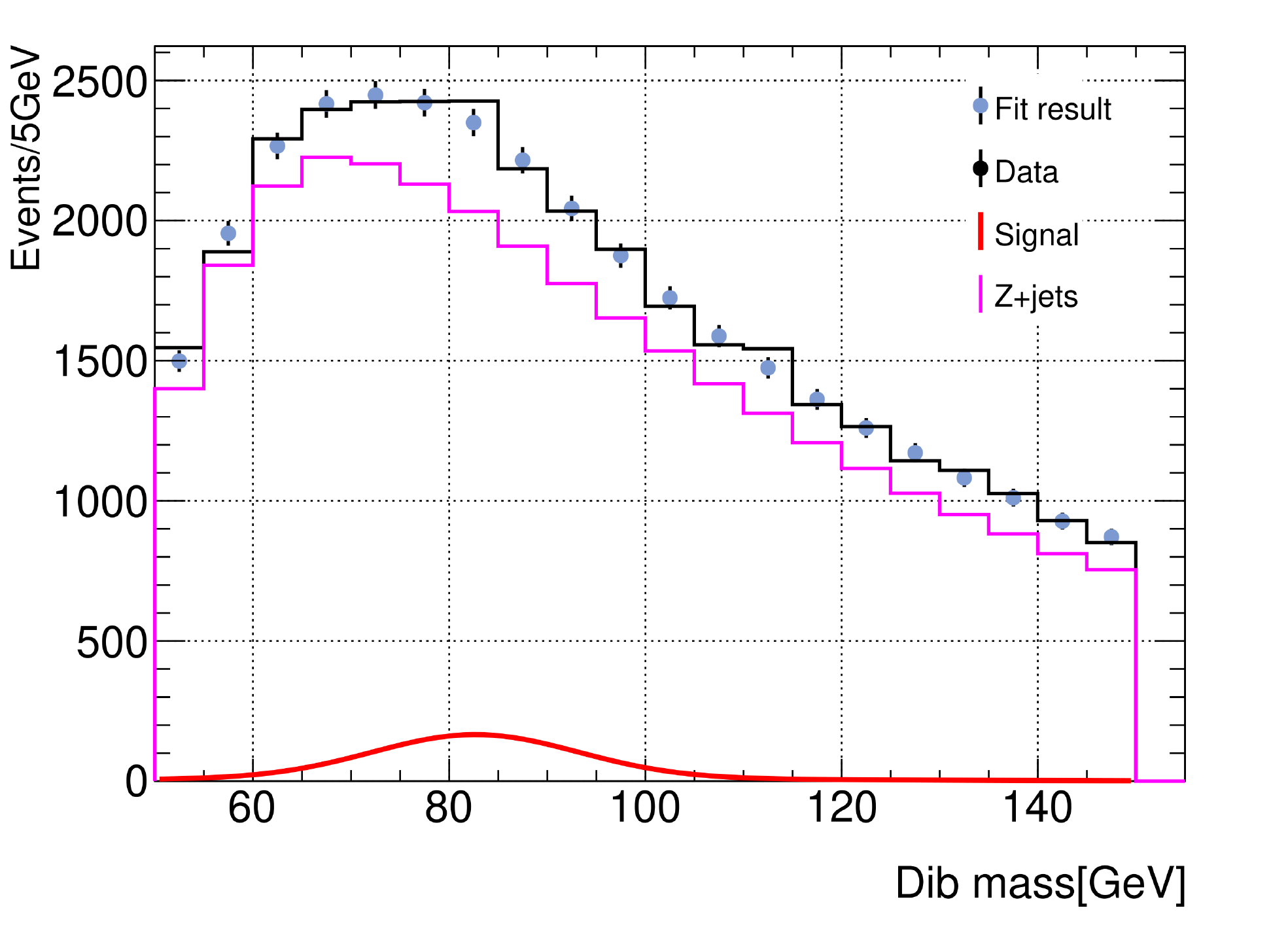} \hspace{3mm}
\caption{Plot of fit to dib-jet mass in data. Musignal is fixed to the expected value of 1.2. }
\label{fig: fig_12}
\end{figure}

Now, the normalisation is fixed at 1.2. It is noted that the error on the mean is correlated to the normalisation: the more signal there is, the easier it is to detect a mass peak. Therefore, a reduction in the normalisation factor translates to an increase in the error on the mean. However, fixing the normalisation gives approximately the right amount of signal and minimises statistical fluctuations due to the background, allowing for the best comparison of the mean between MC and data. Thus, the main conclusions are drawn from the results of this fit, which are presented along with the results from the fake data in Table \ref{tab: tab_2}. A visualisation of the fit is presented in Figure \ref{fig: fig_12}. As can be seen in the plot, the fit seems to model the data reasonably well. This gives another indication that the fit is working well and that the fit results are reasonable. As seen in Table \ref{tab: tab_2}, the mean from this fit and from the MC are compatible. Therefore, there is no obvious indication of interference. However, the error on the mean fitted to data is rather large. One of the reasons for this may be the lower fitted musignal in data, which pushes the error on the mean to higher values, as discussed earlier. 

\subsection{Third fit}
Next, both musignal and the mean are fixed. The $\chi^{2}$ of this fit is 14.1. The fit has 17 degrees of freedom, giving a p-value of 0.65. Indeed, the fit describes the data 15\% better than average when the most help is given to the fit.

\section{Conclusion and Discussion}
The first goal of the project was to experimentally observe Z boson production at low $p_{T}$ through their decay to b quarks, this has been accomplished. Indeed, not only has this peak been identified, the normalisation of the signal is in-line with the expected value. More precisely, 3 sigma evidence for ZZ $\rightarrow$ $b\bar{b}\mu\bar{\mu}$ has been obtained at low $p_{T}$ for the first time. This result is obtained from the returned value of 1.4 $\pm$ 0.4 for musignal in the first fit to data. Another success is the ability of the fit to model the data; as can be seen both visually and by $\chi^{2}$ values, the fit describes the dib-mass distribution sufficiently well.
However, regarding the second goal, the difference between the signal mass peak in data and the signal mass peak in fake data is 0.9 $\pm$ 2.3 GeV - this is compatible with 0 GeV. Therefore, there is not any significant evidence for interference in the $p_{T}$ region explored, or any evidence to suggest that the simulation strategy adopted at the LHC is inadequate.
It is noted, however, that the errors on the signal mass peak in data are large compared to the scale of expected interference. Therefore, the possibility of observing interference remains plausible if errors can be reduced.

There are three main factors which hindered the reduction of the error on the mean, making this project particularly challenging:

\begin{enumerate}
  \item The mass peak in the Z+jets sample is very close to the signal peak, making it hard for a fit to differentiate between the two.
  \item The statistics in the Z+jets sample are poor, with errors five times greater than in data. More events would lower the error. This problem causes difficulties in the modelling of the background, which is predominantly composed of Z+jets. In turn, this makes it rather difficult to test any model and translates to higher errors in the fitted value of the mean.
  \item At low $p_{T}$, the signal to background ratio is considerably worse, compared to regions of higher $p_{T}$.
\end{enumerate}

Considering this project in the wider subset of research in the field of particle physics, this project offers a study of interference at low $p_{T}$ for the very first time. New challenges have been explored and potential solutions have been presented, paving the way for further research in this region.

\section{Further research}
There may be several ways to reduce the error on the mass peak. Potential suggestions will be discussed in this section, harbouring the possibility of further research on interference effects.

Firstly, this study only had access to the dimuon sample. If the dielectron sample could also be accessed, then the number of events could increase by a factor of 2, potentially reducing errors. It is noted that in the dielectron sample, the electrons are triggered on, instead of the muons. There is no theoretical difference between which lepton is used as the trigger.

\begin{figure}[b!]
\centering
\includegraphics[width=77mm]{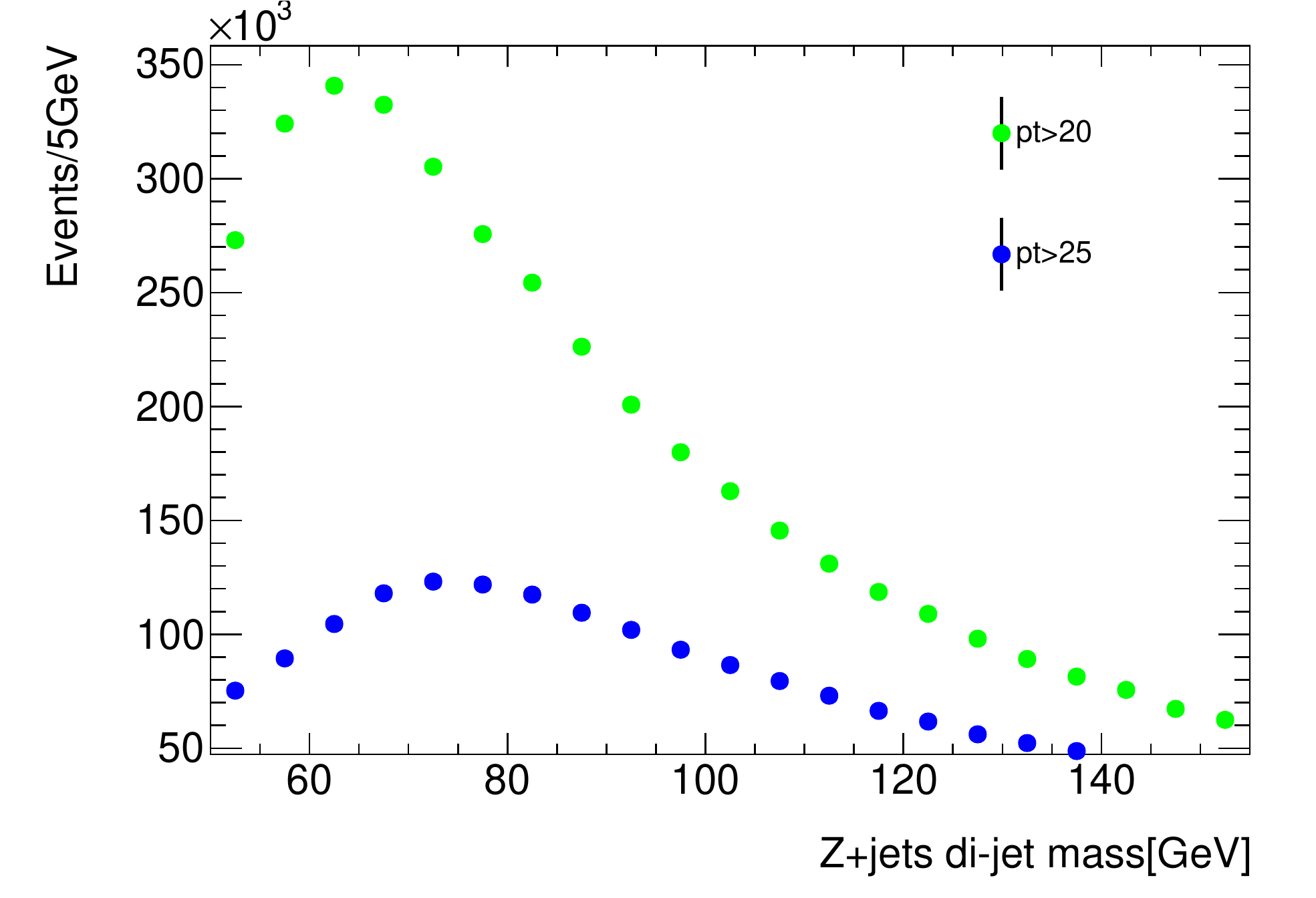}
\includegraphics[width=77mm]{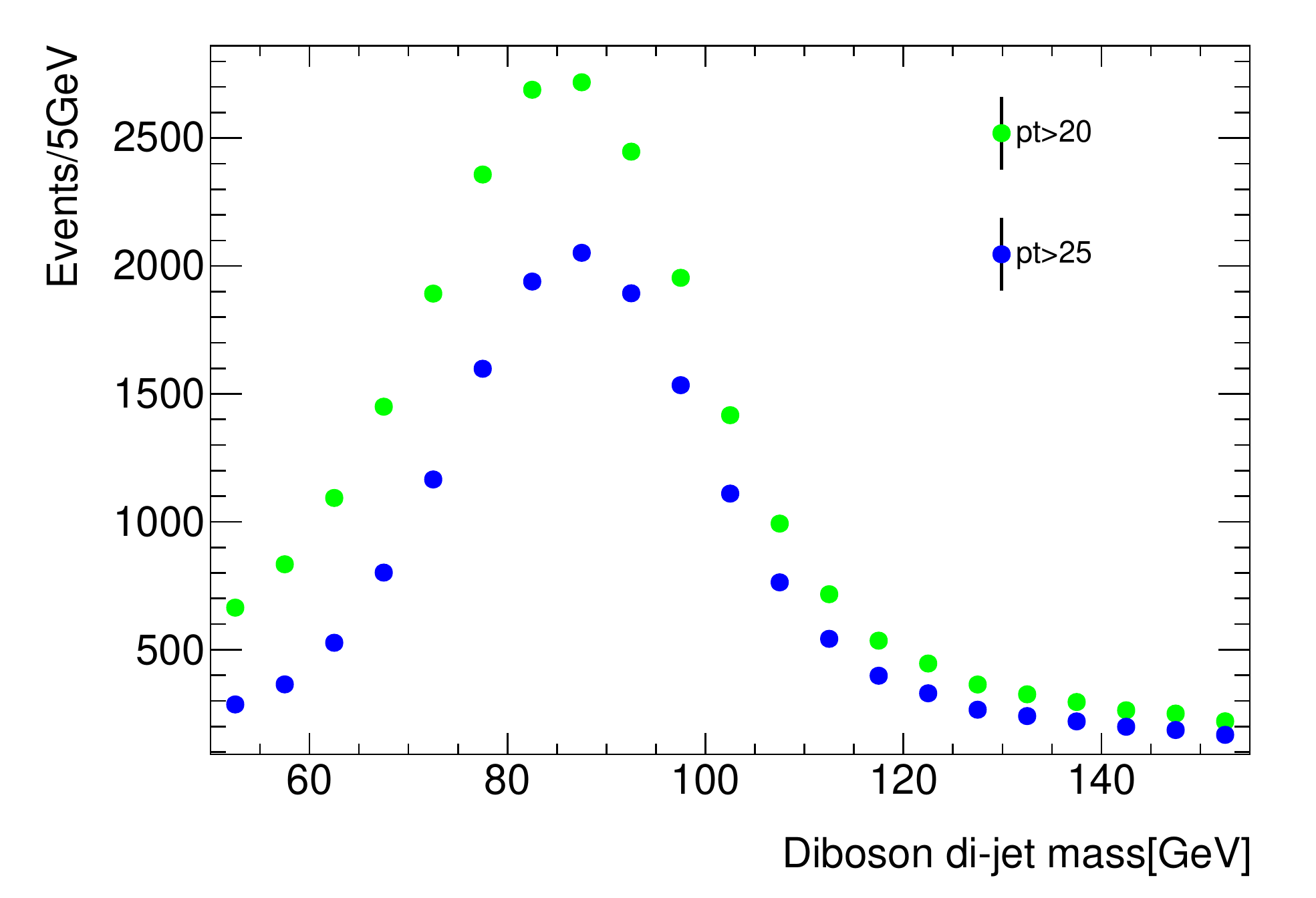}\hspace{3mm}
\caption{Di-jet mass of the Z+jets sample with varied minimum jet $p_{T}$ (left). Peak of the Z+jets is considerably shifted to the lower masses as the minimum jet $p_{T}$ is lowered by 5 GeV. The diboson peak is contained in the same mass region (right).}
\label{fig: fig_13}
\end{figure}

Secondly, this study uses data collected during Run 2, as mentioned earlier. More data will soon become available from Run 3, with 20 times more data being expected to become available by 2040.

Finally, all the events had a minimum $p_{T}$ threshold of 20 GeV on every jet. A plot was drawn of the di-jet mass of the Z+jets at this threshold and the same plot, but with a threshold of 25 GeV, was superposed on the figure. This can be seen in Figure \ref{fig: fig_13}. From the plot, it can seen that the peak of the Z+jets is pushed to lower masses as the threshold is lowered. Based on this, it is hypothesised that if the lower bound of the jet $p_{T}$ can be reduced to 15 GeV, then the peak of the Z+jets would be dropped to even lower masses, increasing the distance between the peak of the Z+jets and the signal peak. If this is the case, then the Z+jets can be modelled as a decaying exponential in the region close to the signal peak, which could in turn make the detection of a signal easier.

\section{Acknowledgements}
I would like to thank my project partner, Francesca Lewis, for working with me to make this project a success. Her insights and contributions were truly invaluable.

\appendix
\section*{Appendix A: MC samples}
Here, the MC simulations will be discussed. In particular, what type of interactions they contain and the physics on which they are based will be explored. 

The Z+jets MC sample has events with a Z $\rightarrow$ $\mu\bar{\mu}$ interaction and at least one jet. Notably, these jets are QCD jets.
Unlike the Z, the W boson does not decay into a pair of charged leptons. Instead, the W can decay into a charged lepton and its neutrino counterpart. As a result, the sample of W+jets is significantly suppressed when two muons are demanded.

\begin{figure}[b!]
\centering
\begin{tikzpicture}
  \begin{feynman}
    \vertex (a) {\(t\)};
    \vertex [right = of a] (b);
    \vertex [above right = of b] (f1)  {\(b\)};
    \vertex [below right = of b] (c);
    \vertex [above right = of c] (f2) {\(\nu_{l}\)};
    \vertex [below right = of c] (f3) {\(l^{+}\)};
    
    \diagram* {
      (a) -- [fermion] (b) -- [fermion] (f1),
      (b) -- [photon, edge label = $W^{+}$] (c) -- [fermion] (f2),
      (c) -- [anti fermion] (f3),
      
    };
  \end{feynman}
\end{tikzpicture}
\hspace{30mm}
\begin{tikzpicture}
  \begin{feynman}
    \vertex (a) {\(q_1\)};
    \vertex [below right = of a] (mid);
    \vertex [above right = of mid] (f1)  {\(q_2\)};
    \vertex [below = of mid] (mid2);
    \vertex [below left = of mid2] (i2) {\(b\)};
    \vertex [below right = of mid2] (f2) {\(t\)};
    
    \diagram* {
      (a) -- [fermion] (mid) -- [fermion] (f1),
      (mid) -- [photon, edge label = $W^{\pm}$] (mid2) -- [fermion] (f2),
      (mid2) -- [anti fermion] (i2),
      
    };
  \end{feynman}
\end{tikzpicture}
\vspace{-2mm}
\caption{Decay of t quark \cite{Paper_3} (left) and t-channel singletop production (right). This channel dominates the sample, although other channels are also in the MC simulations.}
\label{fig: fig_14}
\end{figure}
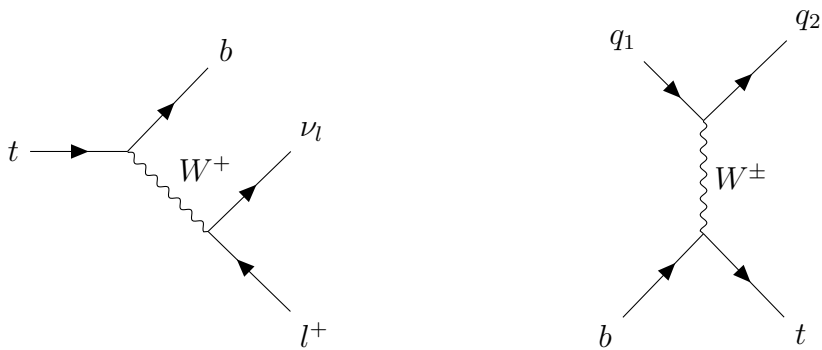

Now, to understand samples containing the top (t) quark, one must understand the decay of the t quark. The t quark is remarkably unique in the family of quarks in that it is significantly heavier than any other quark. For comparison, the second heaviest quark, the b quark, has a mass of 4.5 GeV whereas the mass of the t quark is approximately a staggering 180 GeV \cite{Paper_3}. As a result, the lifetime of t quarks is so short that they decay before forming observable hadrons. This is a remarkable consequence which differentiates the t quark from all the other quarks. Furthermore, the only significant decay mode is t $\rightarrow$  $ bW^{+}$, refer to Figure \ref{fig: fig_14} for the Feynman diagram for this decay. 

Now, the $t\bar{t}$ sample contains events where, rather self-evidently, a $t\bar{t}$ pair is produced, this pair then go on to decay to form b quarks. Also note, the $W^{+}$ bosons from the decay of the t quark can go on to decay into muons, giving the trigger. The singletop sample contains events where one b and one t quark are produced. The t-channel diagram, as shown in Figure \ref{fig: fig_14}, dominates in this sample. The t quark will then follow its normal decay route; note that since there is only one t quark, demanding two muons significantly suppresses the singletop background.

\newpage

\end{document}